\documentclass[3p,singlecolumn,12pt]{elsarticle} % use this to get the feeling of the paper.

\usepackage{amsmath}
\usepackage{amssymb}
\usepackage{lineno}  % Had to put a space before each begin{align} to make linenumber works in paragraphs with equations...
\usepackage{url}                     % For special web formatting.
\usepackage{xcolor}
\usepackage{seqsplit}
\usepackage{datetime}
\usepackage{subcaption}
\usepackage{natbib,hyperref}

\journal{Nuclear Instruments \& Methods in Physics Research}
\begin{document}

\begin{frontmatter}

\title{A High-performance Track Fitter for Use in Ultra-fast Electronics 
%(Time=\currenttime )
}

% \address[Otheraddress]{Institution Name, City, State, Zipcode, Country}
% \author[Otheraddress]{Some. Person}

\address[UBristoladdress]{University of Bristol, Bristol, United Kingdom}
\address[TAMUaddress]{Texas A\&M University, College Station, TX 77843, U.S.A.}
\address[Sahaaddress]{Saha Institute of Nuclear Physics, HBNI, Kolkata, India}
\address[NUaddress]{Northwestern University, Evanston, IL 60208, U.S.A.}
\address[FNALaddress]{Fermi National Accelerator Lab, Batavia, IL 60510, U.S.A.}
\address[UFLaddress]{University of Florida, Gainesville, FL 32611, U.S.A.}
\address[CUaddress]{University of Colorado, Boulder, CO 80309, U.S.A.}
\address[Cornelladdress]{Cornell University, Ithaca, USA}
\address[ULyonaddress]{Universit\'e de Lyon, Universit\'e Claude Bernard Lyon 1, CNRS-IN2P3, Institut de Physique Nucl\'eaire de Lyon, Villeurbanne, France}

\author[UBristoladdress]{E. Clement}
\author[TAMUaddress]{M. De Mattia}
\author[Sahaaddress]{S. Dutta}
\author[TAMUaddress]{R. Eusebi\corref{corauthor}}
\cortext[corauthor]{Corresponding author, +1 (979) 458 7907}
\ead{eusebi@tamu.edu}
\author[NUaddress]{K. Hahn}
\author[FNALaddress]{Z. Hu}
\author[FNALaddress]{S. Jindariani}
\author[UFLaddress]{J. Konigsberg}
\author[FNALaddress]{T. Liu}
\author[UFLaddress]{J. Low}
\author[CUaddress]{R. Patel}
\author[TAMUaddress]{D. Rathjens}
\author[FNALaddress]{L. Ristori}
\author[Cornelladdress]{L. Skinnari}
\author[NUaddress]{M. Trovato}
\author[CUaddress]{K. A. Ulmer}
\author[ULyonaddress]{S. Viret}

%\fntext[cu]{Present address: University of Colorado, Boulder, Colorado, 80309, U.S.A.}

\tnotetext[FPGA]{Field Programmable Gate Array}

\begin{abstract}
This article describes a new charged-particle track fitting algorithm designed for use in high-speed electronics applications such as hardware-based triggers in high-energy physics experiments. 
Following a novel technique designed for fast electronics, the positions of the hits on the detector are transformed before being passed to a linearized track parameter fit.  This transformation results in fitted track parameters with a very linear dependence on the hit positions.
The approach is demonstrated in a representative detector geometry based on the CMS detector at the Large Hadron Collider. The fit is implemented in FPGA\tnoteref{FPGA} chips and optimized for track fitting throughput and obtains excellent track parameter performance. Such an algorithm is potentially useful in any high-speed track-fitting application.
\end{abstract}


\end{frontmatter}

 % \linenumbers

%% main text
% Marco's fitter note is here: /Users/ulmer/Work/doc/Articles/LinearizedTrackFittingNote/
% Stubs should be removed from the text. The reader only needs to hear about hits
% nonlinear should be nonlinear according to 

% -----------------------------------------------------------------------------
\section{Introduction}
\label{sec:intro}

Determining the parameters that describe the trajectory of charged particles traversing a magnetic field is a common problem in high-energy physics applications. There are several solutions that obtain accurate track parameter resolution with methods such as least squares fits utilized in Kalman filters, which employ iterative algorithms to converge toward a desired precision~\cite{FRUHWIRTH1987444, Heinze:2013tma, BUGGE1981365}. In the context of offline processing, where finding accurate parameters takes priority over the computing time and resources spent on it, these powerful algorithms are routinely used in experiments such as CMS and ATLAS.

In the online environment of these high-energy physics experiments, where high-speed electronics are used to make ultra-fast decisions about which particle collision  events to store and which to reject, such sophisticated techniques can easily exceed the limits of available processing time.  For example, the CMS Phase-2 upgraded level-1 trigger~\cite{CMS-TriggerTDR, jorge, tmttelba} aims to find and fit hundreds of tracks within about five microseconds after each proton-proton bunch crossing, which occur every 25\,ns.  Under such stringent conditions, multi-step iterative track fitting algorithms can quickly surpass the allowed time. 

An alternative approach that trades accuracy for reduced computation time is to use a linearized fit. For a track parametrized with $p$ parameters, this type of fit assumes that the true track parameters, \mbox{$p_{\textrm{i}}^{\textrm{true}}, i=1,...,p$,} can be estimated from a linear transformation of the track hit positions, $x_{\textrm{j}}$, where the index $j$ runs over the components of the position of each of the $N$ hits that are part of the fit.

  This is expressed as 
\begin{equation}
p_{\textrm{i}} = \Sigma_{\textrm{j}=1} A_{\textrm{ij}} \delta x_{\textrm{j}} + \overline{p}_{\textrm{i}},
\label{eq:LTF}
\end{equation}
where $A_{\textrm{ji}}$ is a pre-determined set of constants,  $\delta x_{\textrm{j}}$ is defined as $\delta x_{\textrm{j}} = x_{\textrm{j}} - \overline{x}_{\textrm{j}}$ and $\overline{p}_{\textrm{i}}$ and $\overline{x}_{\textrm{j}}$ are the track parameter and hit positions about which the linear expansion is performed. 

% EXPLAIN HOW WE DETERMINE A 
The pre-determined elements of the matrix $A$ are obtained from a sample of tracks representing the true correspondence between track hits and true track parameters. These matrix elements are obtained by minimizing the distance $\textrm{min}(\langle(p_{\textrm{i}}^{\textrm{true}}-p_{\textrm{i}})^2\rangle)$ between the true track parameters $p_{\textrm{i}}^{\textrm{true}}$ and the estimates of Eq.~\ref{eq:LTF} for all track parameters.

In addition, following the principal component analysis described in ~\cite{SVTReviewFermilab} and ~\cite{FTK} it is possible to determine a fit quality parameter that in the limit of validity of the linear approximation is distributed as a $\chi^2$.  This parameter is constructed as the quadrature sum of $N-p$ components, $\chi^2 = \sum_{i=1}^{N-p} \chi_{i}^2$, where each of the components can be expressed as a linear combination of the hit coordinates
\begin{equation}
\chi_i = \Sigma_{\textrm{j}=1}^{N} B_{\textrm{ij}} \delta x_{\textrm{j}},
\label{eq:chi2}
\end{equation}
where the index $i$ runs up from 1 to $N-p$,  and the matrix $B_{\textrm{ij}}$ is a predetermined set of constants obtained following~\cite{FTK} (page 111-112). 
% \begin{equation}
% \begin{pmatrix}
%     p_{1}  \\
%     \vdots \\
%     p_{p}      \\
%     \chi_{i}    \\
%     \vdots \\
%     \chi_{N-p}
% \end{pmatrix}
% =
% \begin{pmatrix}
%     C_{11} & C_{12} & \dots  & C_{1N} \\
%     C_{21} & C_{22} & \dots  & C_{2N} \\
%     \vdots & \vdots & \ddots & \vdots \\
%     C_{N1} & C_{N2} & \dots  & C_{NN}
% \end{pmatrix}
% \begin{pmatrix}
%     x_{1}  \\
%     x_{2} \\
%     \vdots \\
%     x_{N}
% \end{pmatrix}
% \end{equation}

The use of such linear transformations is an especially appealing approach in fast electronics applications since it is based on a small number of addition and multiplication operations that can be computed very quickly and is easily encoded in modern FPGAs. 
A linearized track fit has been previously utilized in the CDF experiment~\cite{cdf} where a hardware track trigger was developed~\cite{SVTReviewFermilab,SVTReview,Adelman:2007zz,SVT-fit}. The approach is also being utilized in the ATLAS experiment~\cite{FTK}. 

However, the assumption of linearity implies that the matrix $A$ is independent of the track parameters, and any such dependence, for example on the track momentum, will introduce nonlinearities. 
In general, the problem of track fitting in most detector geometries involves nonlinearities due to the limited size, the shape, and the orientation of detector sensors, and thus the performance of any linearized fit degrades with the degree of these nonlinearities. In this paper, we describe a new technique that corrects for these nonlinearities and thus improves the performance and overall resource utilization of linearized track fits in hardware. 

The paper is organized as follows. Section~\ref{sec:nl} describes the origin of nonlinearities and the limitations of a linearized track fit using the CMS detector as an example. Section~\ref{sec:corrections} describes a new approach to transform track hit positions, while Section~\ref{sec:performance} shows the performance of this new algorithm. Section~\ref{sec:mem} counts the memory requirements of the technique and Section~\ref{sec:firmware} details the implementation and resource utilization of the algorithm in hardware. Section~\ref{sec:conclusions} presents the conclusions.

% -----------------------------------------------------------------------------
\section{Origin of Nonlinearities}
\label{sec:nl}

The trajectory of a charged particle in a uniform magnetic field can be described by a helix (when neglecting energy loss and other stochastic effects) and is described by five parameters. These are taken to be the charge over the transverse momentum of the particle ($q/p_{\textrm{T}}$), the azimuthal angle of the momentum vector at the point of closest approach to the origin $(\phi_0)$, the transverse impact parameter, defined as the minimum distance between the trajectory and the origin in the transverse plane $(d_0)$, the cotangent of the angle between the momentum vector and the $z$ axis at the point where $d_0$ is evaluated $(\cot\theta)$, and the $z$ coordinate of the trajectory at the point where $d_0$ is evaluated $(z_0)$.

In this note we focus on the use of a linearized fitter with four parameters, where the $d_0$ parameter is assumed to be zero. This assumption is made for simplicity and does not affect the results or the conclusions of this paper. 
%In the following sub-sections we discuss the sources that give rise to nonlinearities in a linearized track fitting.
In the following sub-sections we discuss the sources that give rise to a nonlinear relationship between hit positions and track parameters.

% - - - - - - - - - - - - - - - - - - - - - - - - - - - - - - - - - - - - - - -
\subsection{Nonlinearities from Geometry}
\label{subsec:geo}

We will use the layout of the CMS Phase-2 upgrade detector described in the Technical Proposal~\cite{cms-tp} to illustrate the significant effects played by nonlinearities arising from irregular detector geometry. It should be noted that while the proposed CMS tracker geometry has changed since the writing of this article, the source of nonlinearities in the new geometry are identical to the ones discussed here. In the context of a full level-1 trigger, the track fit technique described in this paper would follow a pattern recognition step, which is not discussed here.

\begin{figure}[!ht]
\begin{center}
\includegraphics[width=\columnwidth]{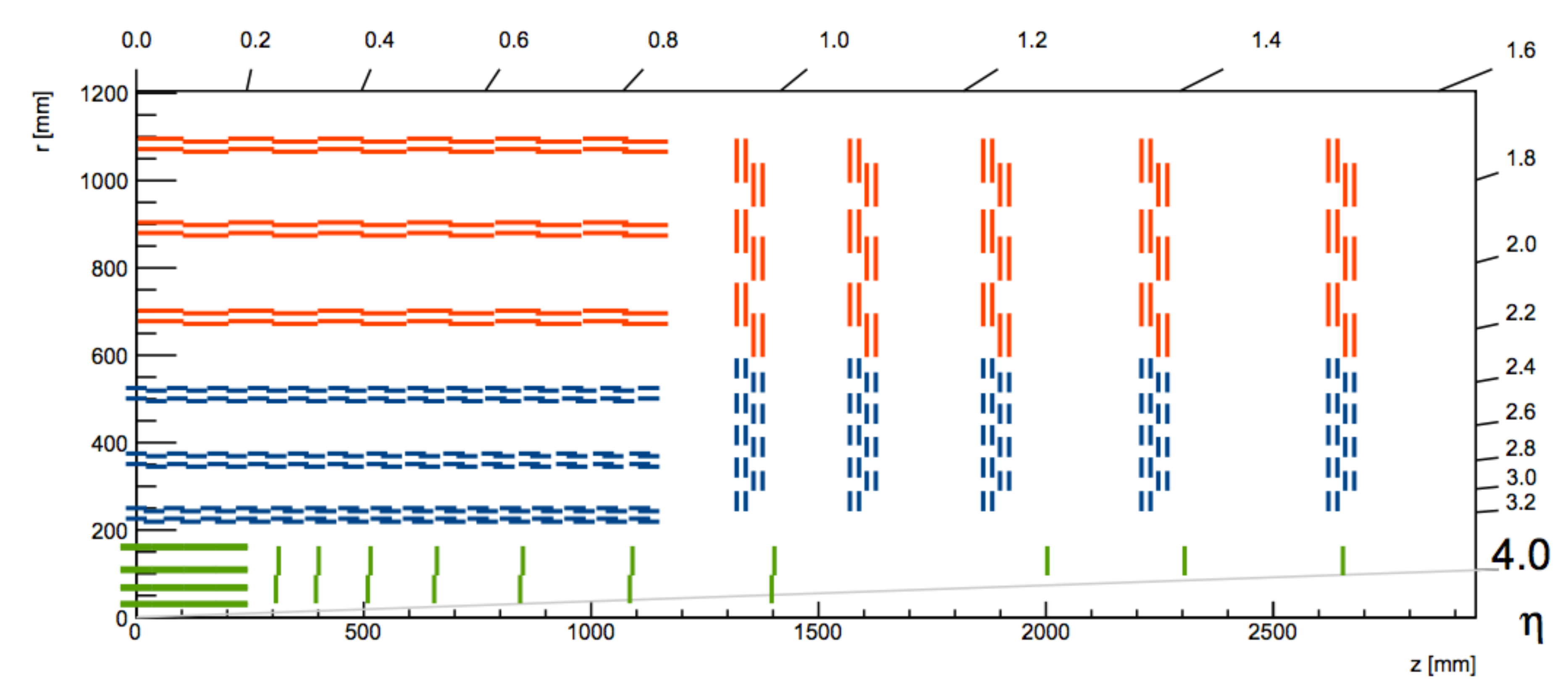}
\caption{Sketch of one quarter of the CMS Phase-2 upgrade silicon tracker as described in the Technical Proposal~\cite{cms-tp}.
The red and blue lines represent silicon strip modules that are used in the track trigger, while the green lines represent silicon pixel modules, which are not considered in the track fit proposed here.}
\label{fig:CMStracker}
\end{center}
\end{figure}

The CMS Phase-2 upgrade detector exemplified in this article is a multipurpose particle detector design with a ``barrel'' consisting of six cylindrical and concentric layers of silicon surrounding the beampipe and two ``endcaps'' consisting of five parallel disks of silicon on the forward and backward ends of the detector, as shown in Fig.~\ref{fig:CMStracker}.  The layers and disks are composed of silicon modules that overlap slightly with neighboring modules to obtain full coverage for charged particles coming from the central luminous region of the detector.

The modules themselves are composed of a pair of silicon sensors separated by about $1-4$~mm to be able to have a local coarse measurement of the track's $p_{\textrm{T}}$ for a given combination of hits in both sensors. This information is read out by the front-end electronics together with the position of the hit in the inner of the double sensors. In this note we do not use or reference that coarse $p_{\textrm{T}}$ estimate, and utilize only the position $(R,\phi,z)$ of the hit in the inner of the double sensors.

% The modules themselves are composed of a pair of silicon sensors separated by about $1-4$~mm to be able to have a rough measurement of the track's $p_{\textrm{T}}$ for a given combination of hits in both sensors. In this note we do not use that rough measurement or reference this feature and utilize only the position $(R,\phi,z)$ of the hit in the inner of the double sensors.  We refer to the position $(R,\phi,z)$ of a signal in the inner of the double sensors in the module as a single hit. 

%The modules themselves are composed of a pair of silicon sensors separated by about $1-4$~mm to be able to have a rough measurement of the track's $p_{\textrm{T}}$ for a given combination of hits in both sensors. The front-end electronics only transmits information on simultaneous hits on both sensors. However, we utilize only the position of the hit on the inner sensor in this note. 

%The modules themselves are composed of double layers of silicon separated by about $1-4$~mm to be able to have a rough measurement of the track's $p_{\textrm{T}}$ for a given combination of hits in both layers. This feature is not exploited or referenced in this note. We refer to the position $(R,\phi,z)$ of a signal in the inner of the double layers as a single hit. 
%The combination of dual hits in a double-layer of modules is called a ``stub'' in CMS's nomenclature but we will refer to the inner module hit with its $(R,\phi,z)$ coordinates of ``stubs'' as hits in this note.

The inner modules (drawn in blue in Fig.~\ref{fig:CMStracker}) consist of a macro-pixel sensor and a strip sensor on top of each other. The macropixels and strips are $1.5\,$mm and $2.4\,$cm long in the $z$ direction, respectively. The pitch is $100\,\mu$m for both sensor types.
The outer modules (drawn in red) have parallel silicon strips of 90\,$\mu$m pitch measuring 50\,mm in length along the $z$ direction. 

We use the official right-handed coordinate system of CMS in which the origin lies at the geometrical center of the detector, the $z$ axis lies longitudinally along the detector and along the direction of the magnetic field, and the $x$ and $y$ axes form the transverse plane.  Given the cylindrical symmetry of the detector and its internal magnetic field we also use cylindrical coordinates ($R$, $\phi, z$) in this note, where $R$ and $\phi$ are constructed from the $x$ and $y$ coordinates of the official CMS coordinate system.

For a perfectly cylindrical detector, the radial coordinates do not provide information useful for a principal component analysis of the hit correlations since 
they are at fixed values for each layer.
In this perfect detector it is sufficient to consider the hit position information ${(\phi,z)}$ from different layers as separate inputs. This approximation is almost always used because it reduces the number of constants in the matrix and the resulting number of operations needed to obtain the result, making the matrix multiplication faster and less memory intensive.

\begin{figure*}[!ht]
 \begin{center}
  \includegraphics[width=0.34\textwidth]{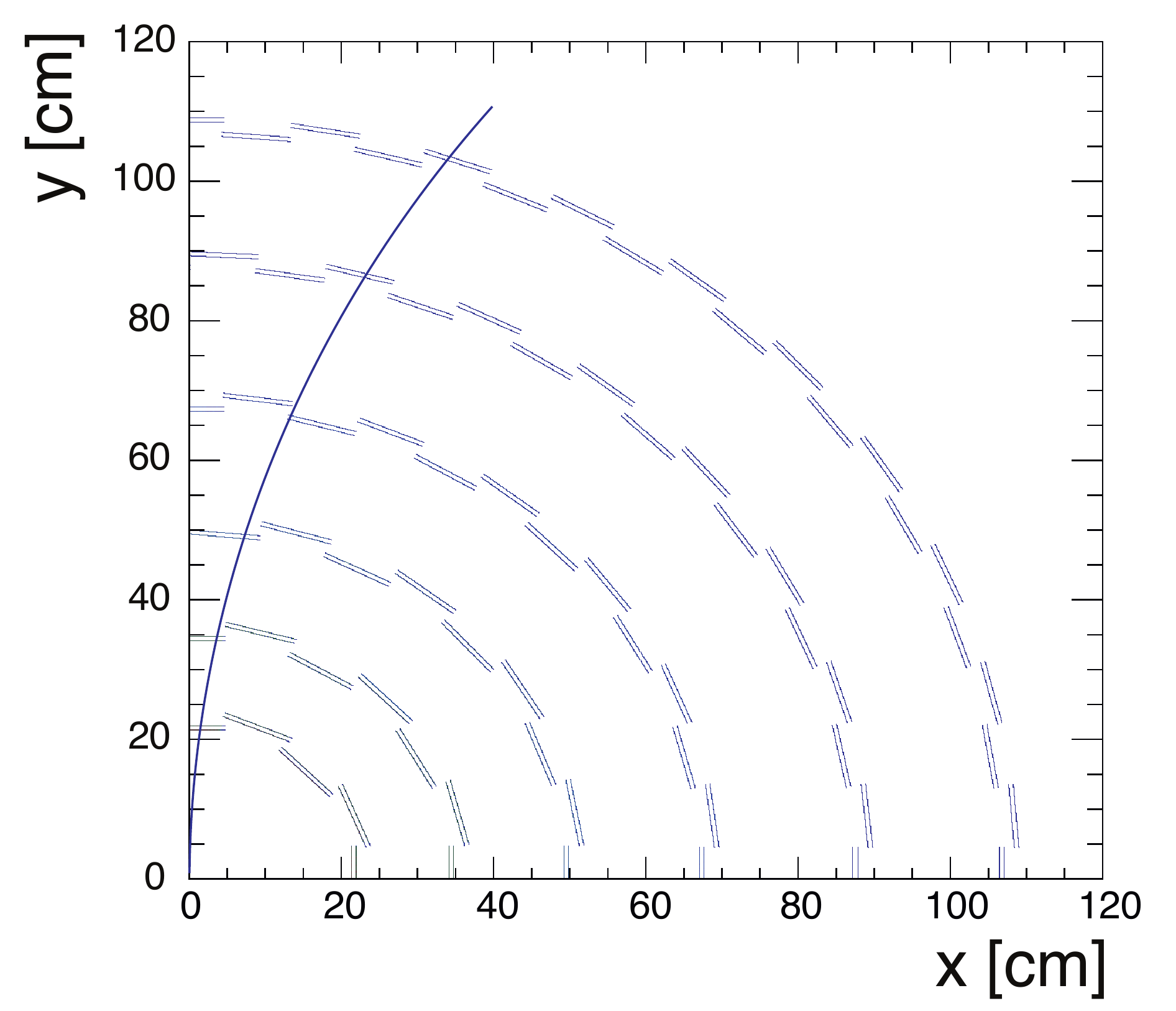}
  \includegraphics[width=0.315\textwidth]{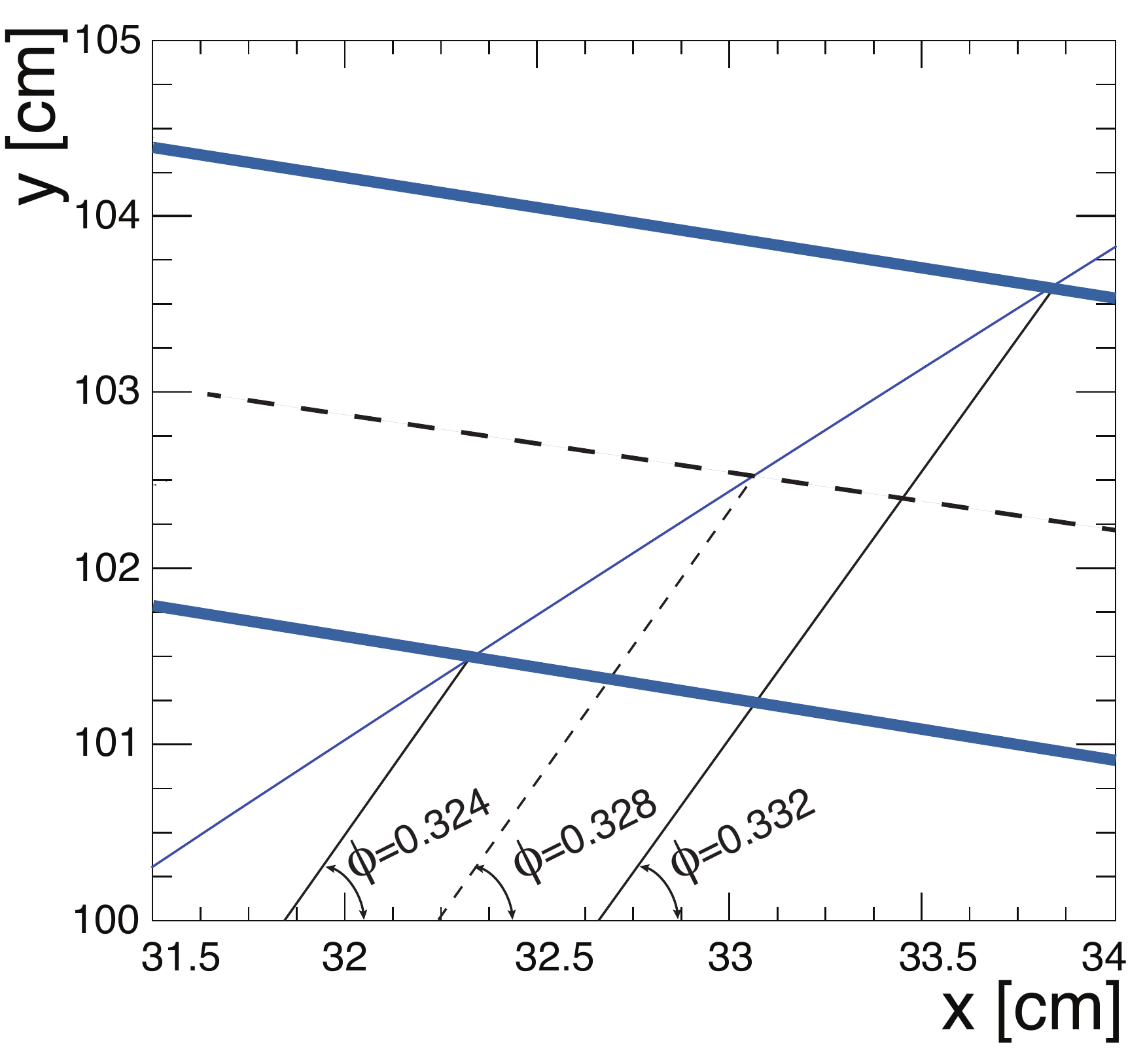}
  \includegraphics[width=0.32\textwidth]{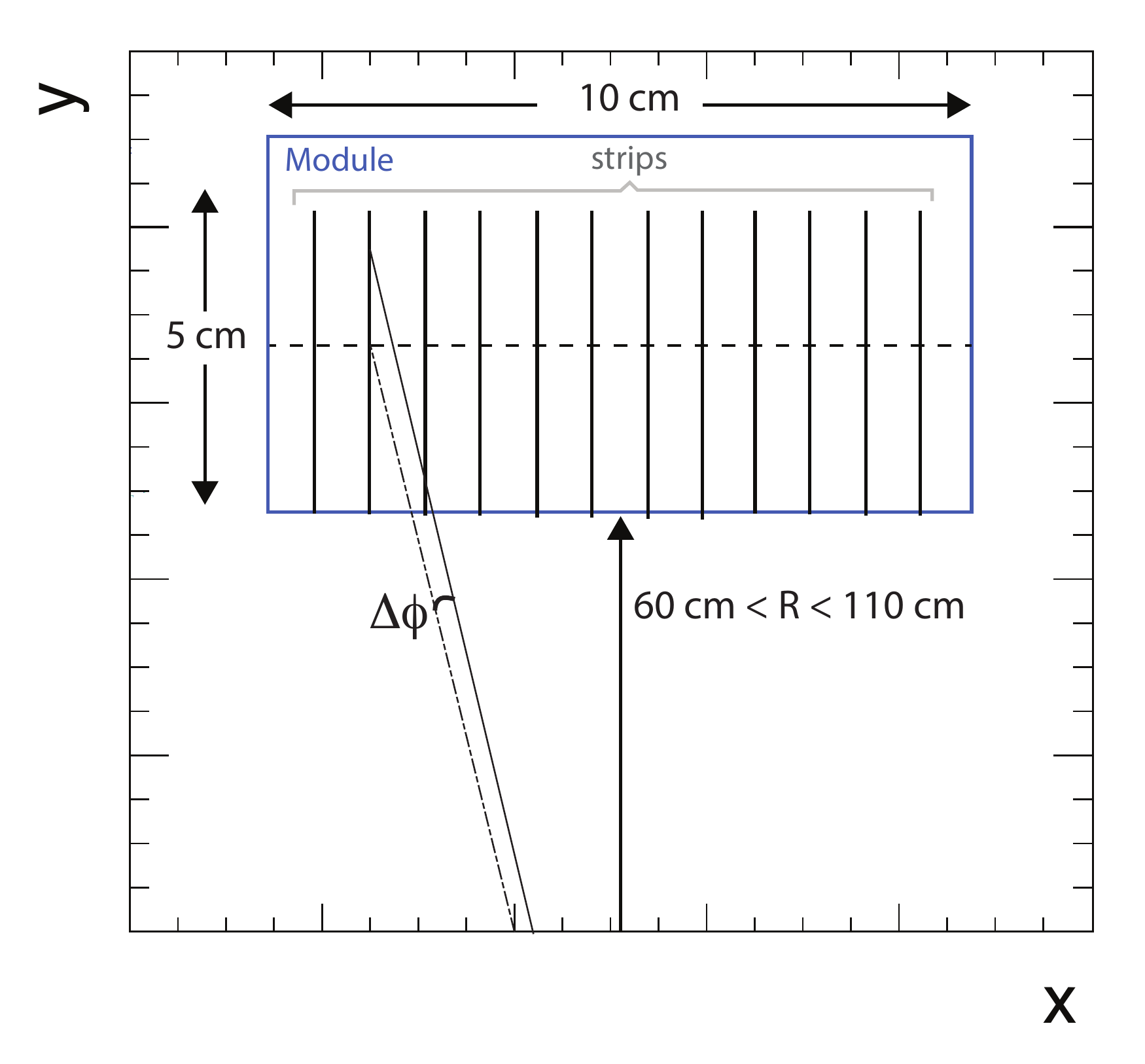}
 \caption{Geometrical origin of nonlinearities. \textbf{Left:} staggering and flatness of the modules in the barrel as indicated by the positions of reconstructed hits from simulated single muon events. The line depicts a $2\,$GeV track. \textbf  {Center:} zoomed view of the left figure in the area of the tracks crossing the outermost layer.  The thick lines represent the average radius of the innermost and outermost modules in the layer, while the dotted line shows the average radius of all modules in the layer. The angles are the $\phi$ coordinates of the hits that would be registered for this track depending where the hit is actually produced. \textbf  {Right:} Parallel-strip modules in the disk. Sketch of a 2D module in the disks of the endcap region and some of the strips within. The figure shows the difference in angle of the $\phi$ coordinate due to the parallel arrangement of the strips.}
 \label{fig:NonLin}
 \end{center}
\end{figure*}

In a real detector, however, the hit coordinates for a given layer are not at the same radius because of two effects: the modules are
flat, and they are staggered to allow for overlaps and complete coverage with active material. This is shown in Fig.~\ref{fig:NonLin} (left) where the barrel layers are depicted and pairs of modules at the same angle $\phi$ correspond to different $R$ positions and have significant variation in radius.

The staggering of the modules results in nonlinearities in the transformation from tracker hits to track parameters. As an example, Fig.~\ref{fig:NonLin} (center) illustrates the size of these nonlinearities in the outermost layer of the detector for a $p_{\textrm{T}}=2\,$GeV track. The assumption that the modules in that layer are at a constant radius introduces a $p_{\textrm{T}}$-dependent misestimation in the $\phi$ coordinate of the hit of about four mrad in either direction.

Likewise, an assumption of a constant radius introduces a change in the estimated $p_{\textrm{T}}$ track parameter. 
To estimate the change in $p_{\textrm{T}}$ that would compensate for the shift in radius, we consider a track assumed to come from the center of the detector in the transverse plane, that is with $d_0 = 0$. In that case we can write the coordinates $\phi$ and $z$ of a point on the trajectory as a function of its radius $R$ and the track parameters $q/p_{\textrm{T}}$, $\phi_0$, $cot(\theta)$, and $z_0$ as

\begin{align}
\phi &= \phi_0 - \arcsin\left(\frac{R}{2\rho}\right)\,,\textrm{ and} \label{eq:trajectoryPhi} \\
z &= z_0 + 2\rho \arcsin\left(\frac{R}{2\rho}\right)\cot\theta\,, 
\label{eq:trajectoryZ}
\end{align}
where $\rho = 0.33\,\textrm{m} \cdot  (p_{\textrm{T}}/\textrm{GeV})/(B/\textrm{T})$ is the radius of curvature of the trajectory in the transverse plane and is directly proportional to the $p_{\textrm{T}}$ of the track and inversely proportional to the magnetic field $B$.
% , where we note that the description here ignores stochastic effects such as multiple scattering. 

From Eq.~\ref{eq:trajectoryPhi} we can compute the $p_{\textrm{T}}$ of a track that would produce, on the outermost staggered module, a hit with the same $\phi$ of the hit produced by the $2\,$GeV track in the innermost module. This relative variation in $p_{\textrm{T}}$ is approximately independent of $p_{\textrm{T}}$ and is approximately inversely proportional to the variation of the radius of the hit. The biggest variation in $p_{\textrm{T}}$ is seen when considering the innermost layer where it can be as big as about $11\%$, while in the outermost layer it is the smallest and it is about 2.4\%.
These effects significantly degrade the online performance and need to be addressed to reach the typical offline track $p_{\textrm{T}}$ resolution of around $0.5\%$ ($2\%$) for 2 (100) GeV tracks~\cite{CMS-TrackerTDR}.

A second effect that causes biases and reduces the precision of the measured parameters is the planar shape of the modules, which causes a variation of the radius of the hit within a single module. This effect is most important for the innermost layer, where 16 modules cover the full $\phi$ angle with each module covering approximately 0.4 radians. Assuming the center of the module to be its closest point to the origin and having a radius of approximately 23\,cm, the radius at the edge is approximately 23.5\,cm, a variation of about $2\%$. For a $2\,$GeV track this translates into a similar effect on the $p_{\textrm{T}}$ of about $2\%$. For the second innermost layer this effect is reduced to less than $1\%$ by the increase in the number of modules (24) and it is reduced further as the number of modules per layer increases at even larger radii.

Furthermore, Eq.~\ref{eq:trajectoryZ} shows that the effect of the variation of the radius on  $\cot\theta$ is similar to the effect on $p_{\textrm{T}}$ in the case of the $\phi$ coordinates.  The effect is maximal in the outermost layer where it is about $11\%$ and minimum in the innermost layer where it is about 2.5\%.  Again, this is a significant effect in the performance of a linearized fit.

A third source of nonlinearities from detector geometry stems from the parallel orientation of the strips on the silicon module. This is problematic in the endcap disks, where the outer modules are arranged so that their $\sim 5\,$cm strips roughly point towards the $z$ axis. However, since the strips are parallel to each other, if one strip points straight to the $z$ axis, all others will not. Figure \ref{fig:NonLin} (right) shows the geometry of an outer module in the disk as seen in the transverse plane. Since the outer modules of the disk do not provide information on the position of the charge deposit along the strip, each hit on each strip in the disks is assigned the $R$ coordinate of the center of the strip drawn as a dashed line in the figure. 
The figure shows how the parallel strips arrangement causes the correct radius $R'$ of the hit to deviate from the assigned radius $R$ if the particle did not pass through the center of the strip. As a consequence, the $\phi$ coordinate of the hit is also biased, as shown in the figure. This source of nonlinearity does not exist in the barrel, where the strips are directed along the $z$ axis. It would also disappear from the disks if either the strips were arranged in a radial fashion all pointing toward the $z$ axis or if a more accurate  position of the hit along the strip was provided, as happens in the inner modules. 
%
%
% - - - - - - - - - - - - - - - - - - - - - - - - - - - - - - - - - - - - - - -
\subsection{Nonlinearities from Functional Form}
\label{subsec:form}
Another source of nonlinearities comes from the track equations themselves. Equation \ref{eq:trajectoryPhi} shows that the equations that describe the trajectory of tracks are nonlinear and that simply using linearized equations will not give the optimal performance. It should be noted that other effects, such as a realistic magnetic field that is not fully homogeneous over the tracking volume, will increase the complexity of the functional form of Eq.~\ref{eq:trajectoryPhi}.

While this source of nonlinearity seems irreducible in the usage of a linearized track algorithm, we will show in the following sections how it can be circumvented.

% - - - - - - - - - - - - - - - - - - - - - - - - - - - - - - - - - - - - - - -
\subsection{Standard Mitigating Mechanisms}
\label{subsec:coping}

To mitigate these effects the most common approach taken by the high-energy physics field is to generate many different sets of constants, each to be applied in a smaller region of the detector. By reducing the region of application in the detector the effects of the nonlinearities in that region are effectively reduced.  The cost of such an approach is the need to generate, store, and then quickly retrieve a large number of sets of constants and apply each to the correct region of the detector. The constants must also be stored in the hardware and be accessible to the fit, therefore limiting the amount of memory available for the fitter engine itself. A tradeoff must be made between the resolution of the fitted parameters, which depends on the size of the region of application, and the total number of constants needed. The typical number of regions used in high-energy physics experiments ranges from a few hundred thousand to several million. 

Furthermore, it should be noted that the nonlinearities, which are the root-cause of the lack of performance of any linearized track fitter, are simply not addressed. In this approach, the underlying nonlinearities remain in place reducing the performance even within the small regions; it is just the interval of application of each set of constants that is reduced.
%
% -----------------------------------------------------------------------------
\section{Hit-position Transformations}
\label{sec:corrections}
Instead of reducing the region of application of each set of constants while increasing the total number of regions, we describe here an approach that directly addresses the nonlinearities of the system, allowing for a considerably more performant track fitter for a given region of application, or alternatively allowing the region of application to be extended significantly for a given desired performance.

Our approach starts by restoring the cylindrical symmetry at the input stage of the linearized fitter by transforming the locations of the hits on a physical layer to new positions that lie on an ideal cylindrical layer of fixed radius, $R'$. In addition, the transformation is made such that the coordinates of the new hit position $(R',\phi',z')$ have a linear functional dependence on the track parameters:
\begin{align}
\phi' &= \phi_0 - \frac{R'}{2\rho} \label{eq:phiprime},  \\
z' &= z_0 + R' \cot\theta\, \label{eq:zprime} .
\end{align}
These linear parameterizations are chosen because they are the first order approximations of Eqs.~\ref{eq:trajectoryPhi} and~\ref{eq:trajectoryZ}. Using Eqs.~\ref{eq:trajectoryPhi} and \ref{eq:trajectoryZ}, and Eqs.~\ref{eq:phiprime} and~\ref{eq:zprime}, we write the hit coordinates on the ideal layer $(R',\phi',z')$ in relationship to the coordinates on the physical layer $(R,\phi,z)$ and the track parameters as: 
\begin{align}
\phi' & =  \phi - \frac{R'}{2\rho} + \arcsin\left(\frac{R}{2\rho}\right) \label{eq:corPhi},\\
z' & = z + 2\rho \cot(\theta)\left[\frac{R'}{2\rho} - \arcsin\left(\frac{R}{2\rho}\right)  \right]  \label{eq:corZ} .
\end{align}

Let's consider the first order approximation of Eq.~\ref{eq:corPhi} around $R/(2\rho) = 0$:
\begin{equation}
\phi' = \phi + \frac{R - R'}{2\rho}.
\end{equation}

The size of the shift in the $\phi$ hit coordinate required to extrapolate to the ideal layer is $s=\phi'-\phi=(R-R')/2\rho$. However, we note that
the relative accuracy of the shift, $\delta s/s$, is directly proportional to the relative accuracy of the radius of curvature, which by construction is also that of the transverse momentum:
\begin{equation}
\frac{\delta s}{s} = \frac{\delta \rho}{\rho} =  \frac{\delta p_{\textrm{T}}}{p_{\textrm{T}}}.
\end{equation}
This shows that an a-priori value, or ``pre-estimate'' of the transverse momentum with a large relative resolution can be used to evaluate the shift in the $\phi$ coordinates to that same relative precision. For example, a $p_{\textrm{T}}$ pre-estimate with a $3\%$ relative resolution can produce a first-order correction that is only $3\%$ off from ideal.  Our technique develops directly from this fact.

We start by performing a fast and coarse pre-estimate of the track quantities $q/(2\rho)$ and $\cot\theta$. These pre-estimates are also performed as a linear fit with a single set of constants determined from averaging over the entire region of application. As described in Section~\ref{subsec:geo} these pre-estimates have sub-optimal resolution on the track parameters providing a transverse momentum relative resolution of $\sim 2$--$3\%$. While this coarse measurement is worse than the ultimately desired precision on the track parameters, it will allow the transformation of the projected $\phi$ position to the same relative precision.

We then transform the hit positions to the ideal layer. To minimize processing time in real hardware applications instead of directly using Eqs.~\ref{eq:corPhi} and \ref{eq:corZ} we utilize the following polynomial approximations:

\begin{align}
\phi' &= \phi + \frac{R - R'}{2\rho} + \frac{1}{6} \left(\frac{R}{2\rho}\right)^3 \label{eq:actualcorPhi},\\
z' & = z - \cot(\theta) (R-R')  - \frac{\cot(\theta) R^3}{6 (2\rho)^2}\label{eq:actualcorZ}. 
\end{align}
 % seqsplit{https://indico.cern.ch/event/591143/contributions/2385614/attachments/1382277/2106149/TTAM\_TrackFit\_12\_8\_16.pdf\#search=De\%20Mattia}

\begin{figure}[!htb]
\begin{center}
\includegraphics[width=\columnwidth,clip]{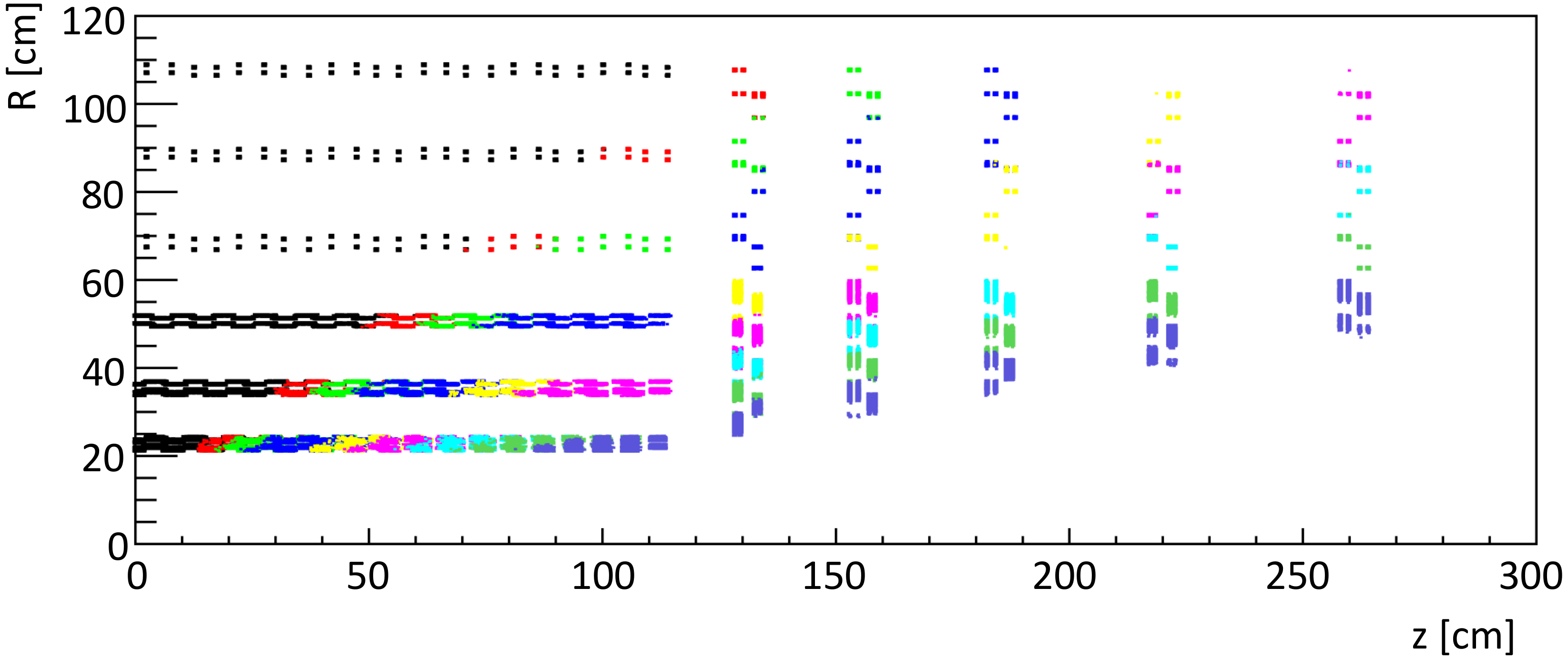}
\includegraphics[width=\columnwidth,clip]{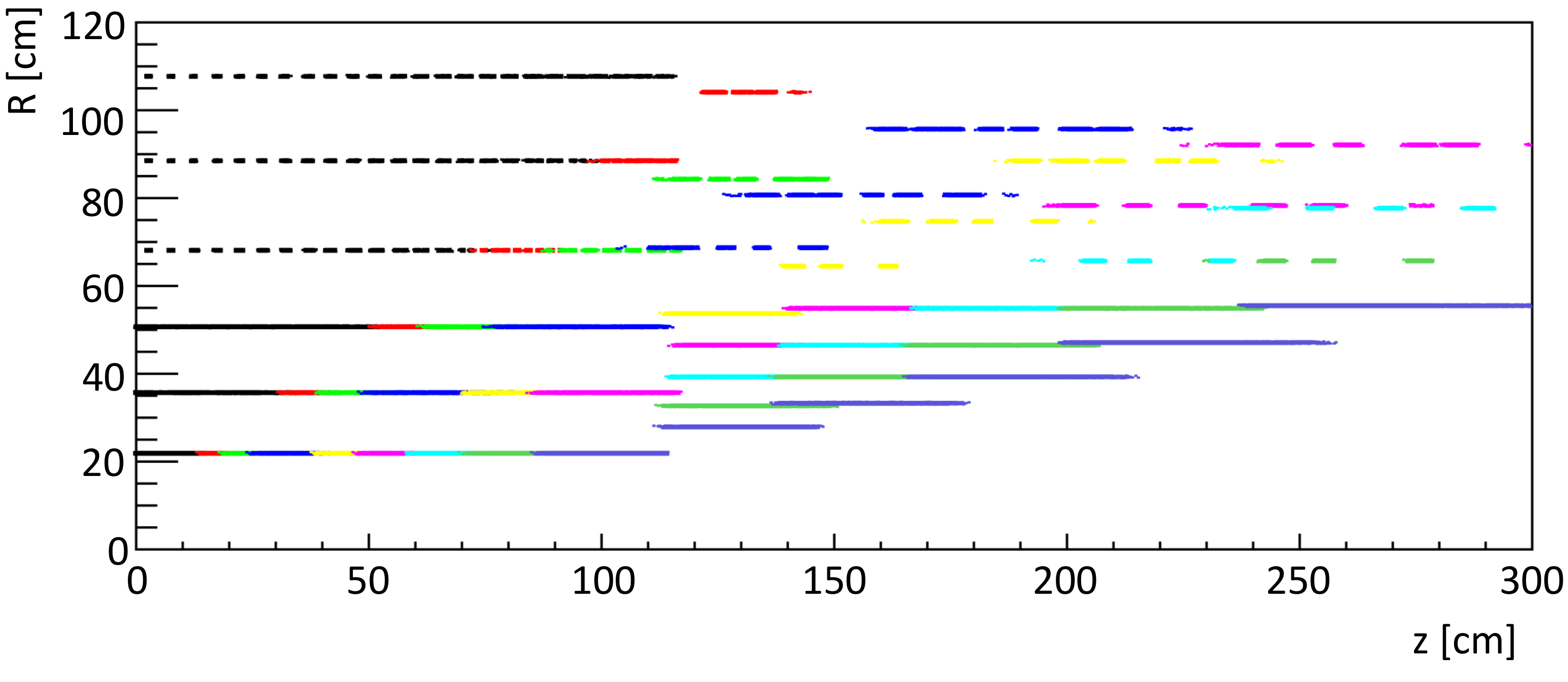}
\caption{Hit positions in the ($r$, $z$) plane. The colors represent hits assigned to the different regions. Top: original positions. Bottom: transformed hit coordinates. }
\label{fig:projections}
\end{center}
\end{figure}

The result of these transformations is illustrated in Fig.~\ref{fig:projections} where the top panel shows the original hit locations and the bottom panel shows those hits transformed onto common cylindrical surfaces. The (positive $z$) half of the CMS detector has been divided into 14 regions. The plot shows the nine major geometrical regions depending on the combination of modules in the barrel layers and disks, which are shown in different colors in the figure.
Five of these regions are further separated in two but are left out of this figure for visualization purposes.
% Marco Wrote Other smaller regions used to fill gaps due to the $z$ spread of the track vertices are not easily visualized in this projection and are left out for visualization purposes.
  Notice how the whole barrel layer six constitutes a single region and how the hits in the disks were also extrapolated to ideal cylindrical layers in each individual region.

We can also use the pre-estimated quantities to reduce the nonlinearities coming from having parallel-strip arrangements in the outer modules of the disks as discussed at the end of Section~\ref{subsec:geo}. For those modules located at a nominal radius $R$ we estimate the correct radial value of the hit coordinate, $R_{\textrm{ex}}$, as a linear extrapolation from the hit $z$ coordinate in the outermost barrel layer or the outermost inner disk as:

\begin{align}
R_{\textrm{ex}} = (z-z_0) \tan\theta \label{eq:Rextra}.
\end{align}
We then correct further the $\phi$ coordinate of hits in these modules as:

\begin{align}
\Delta\phi = p\cdot(h_{\textrm{sn}} - m_{\textrm{sn}}) \cdot \frac{ R_{\textrm{ex}} - R }{R^2}  \label{eq:actualNonRadial},
\end{align}
where $h_{\textrm{sn}}$ is the strip number of the actual hit, $m_{\textrm{sn}}$ is the strip number of the center strip in the module, and $p$ is the distance separating any two strips in the module. Note that the computation of $R_{\textrm{ex}}$ requires a pre-estimate value of $\tan\theta$, rather than $\cot\theta$. To avoid performing a costly division in the hardware, we simply compute $\tan\theta$ as another pre-estimate quantity together with $q/(2\rho)$ and $\cot\theta$. 

It is important to notice that this technique transforms the individual position of a hit to a cylindrical geometry using only a track pre-estimate and the individual hit position. The technique does not use any information from other hits and is therefore blind to the arrangement of hits that stems from a given detector geometry. That is, since it will perform an intermediate transformation that changes that geometry into an ideally cylindrical barrel, this technique can be deployed in any detector geometry. 

A second linearized track fit is then performed on the transformed hit coordinates, in each region, which is used to obtain the final track parameters. This technique directly addresses the geometrical
nonlinearities described in Section~\ref{subsec:geo}. In addition,  Eqs.~\ref{eq:corPhi} and \ref{eq:corZ} directly target the nonlinearities arising from the functional forms
described in Section~\ref{subsec:form}. This technique mitigates both sources of nonlinearities. The net result is a least-squares fit to the track hits where the transformed
coordinates have been used to significantly reduce the nonlinearities based on the pre-estimate fit.

% -----------------------------------------------------------------------------
\section{Performance of a Track Fitter Based on Transformed Hits}
\label{sec:performance}

The approach of projecting hit positions onto idealized layers before a linearized track fit has been explored using the CMS Phase-2 upgrade tracker geometry. Details of the implementation of this proof of principle and the track fitting results are presented in this section.

A sample of simulated single muons, generated evenly in $1/p_{\textrm{T}}$ between 0.5/GeV and 0.005/GeV, is used to determine the optimal coefficients needed in each region to implement this technique. In general, however, any set of good tracks can be used, and a sample of tracks reconstructed directly in data would allow for detector misalignment effects to be taken into account directly. 
The sample of muons was simulated using the official CMS Monte Carlo simulation described in~\cite{CMS-TrackerTDR} using the official CMS software package CSSMW~\cite{CMSSW}.

%-Use clean single muon events that leave six hits in the detector. (Delta ray cleaning? 7 stub cases?)
% https://indico.cern.ch/event/591143/contributions/2385614/attachments/1382277/2106149/TTAM_TrackFit_12_8_16.pdf#search=De%20Mattia

In the case of the specific example of the CMS detector discussed here the tracker is divided into a total of 14 different regions which correspond to different combinations of barrel and endcap detector modules. Tracks that pass entirely through the barrel are considered in one region, tracks that pass through five barrel layers and one endcap disk are considered in another region, etc. Additionally, regions containing at least one endcap disk are further split in two to consider the different characteristics of modules in the inner and outer regions of the detector. 
For each region, an individual set of constants is determined. Within each region, the entire detector utilizes the same set of constants for all $\phi$ coordinates. To further suppress nonlinearities at low $p_{\textrm{T}}$, different sets of constants are trained for tracks above and below $p_{\textrm{T}} = 10$\,GeV and the correct set of constants is determined on the fly based on the $p_{\textrm{T}}$ pre-estimate. 

In the CMS geometry layers have overlapping modules allowing a single track to produce hits in several modules of a single layer. Since our technique uses a single hit per layer (as discussed in Section~\ref{subsec:geo}) we mask off fixed regions in the outer modules where hits are likely to be coming from tracks that also leave hits on an inner, partly overlapping, module. This results in having almost exclusively one hit per layer at a negligible efficiency loss. 

Tracks in this geometry produce hits in up to seven different layers or disks, with the typical case producing hits in six layers.  In this study we consider tracks with one hit in five or six different layers. Tracks with hits in seven different layers are reduced to six by removing the hit in the outermost layer.

\begin{figure}[!ht]
\begin{center}
\includegraphics[width=0.48\textwidth,height=4.8cm]{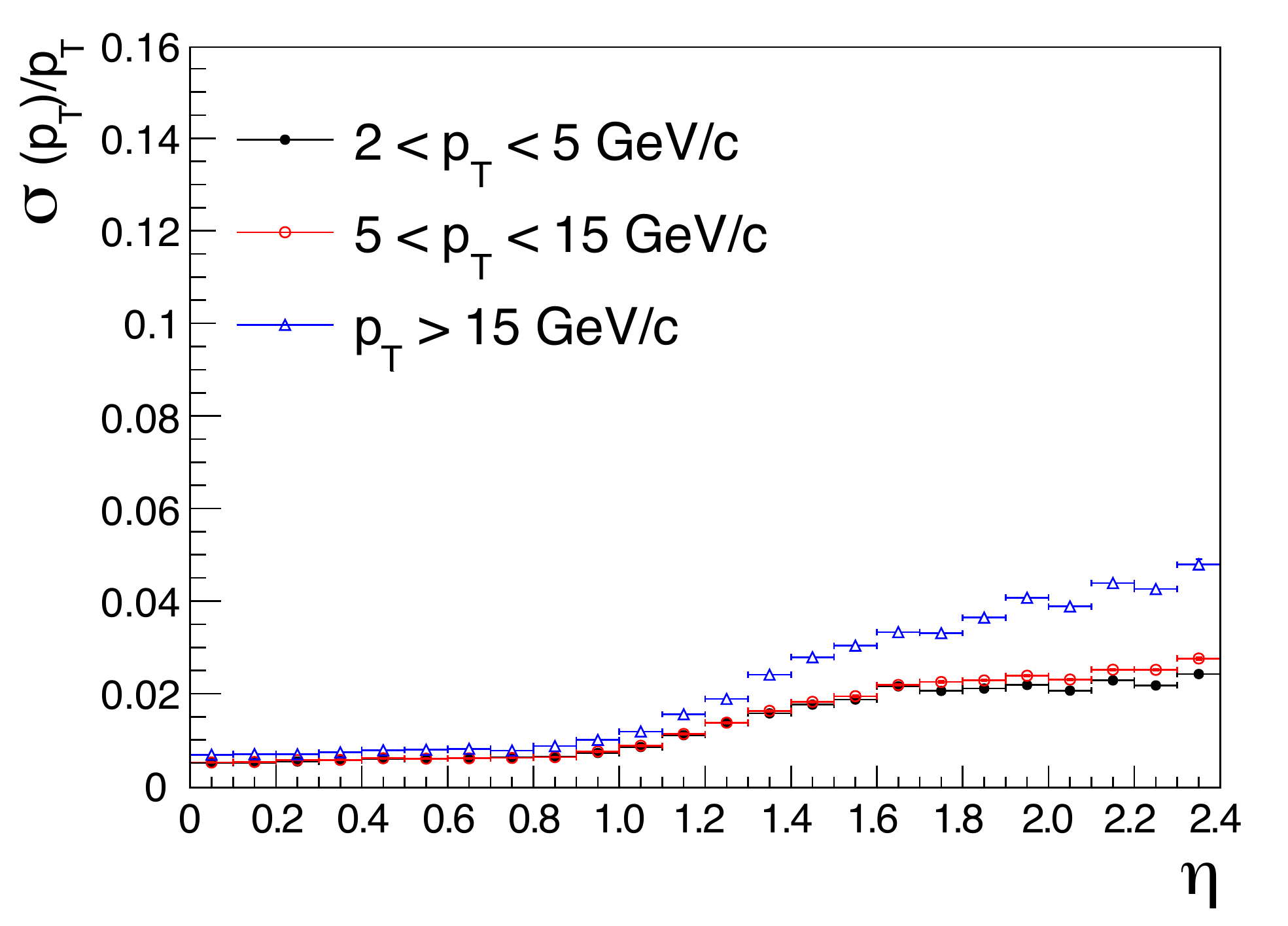}
\includegraphics[width=0.48\textwidth,height=4.8cm]{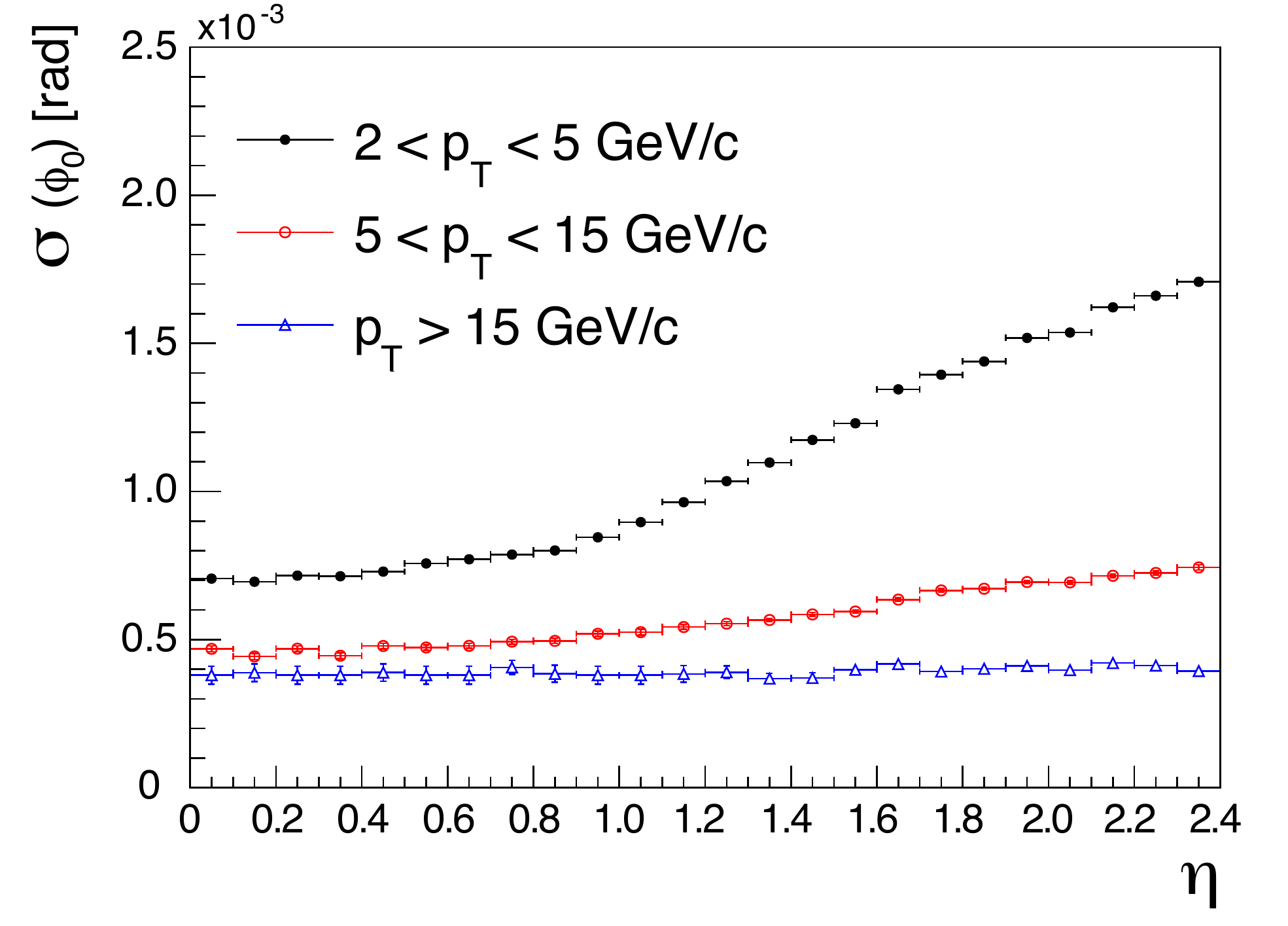}
\includegraphics[width=0.48\textwidth,height=4.8cm]{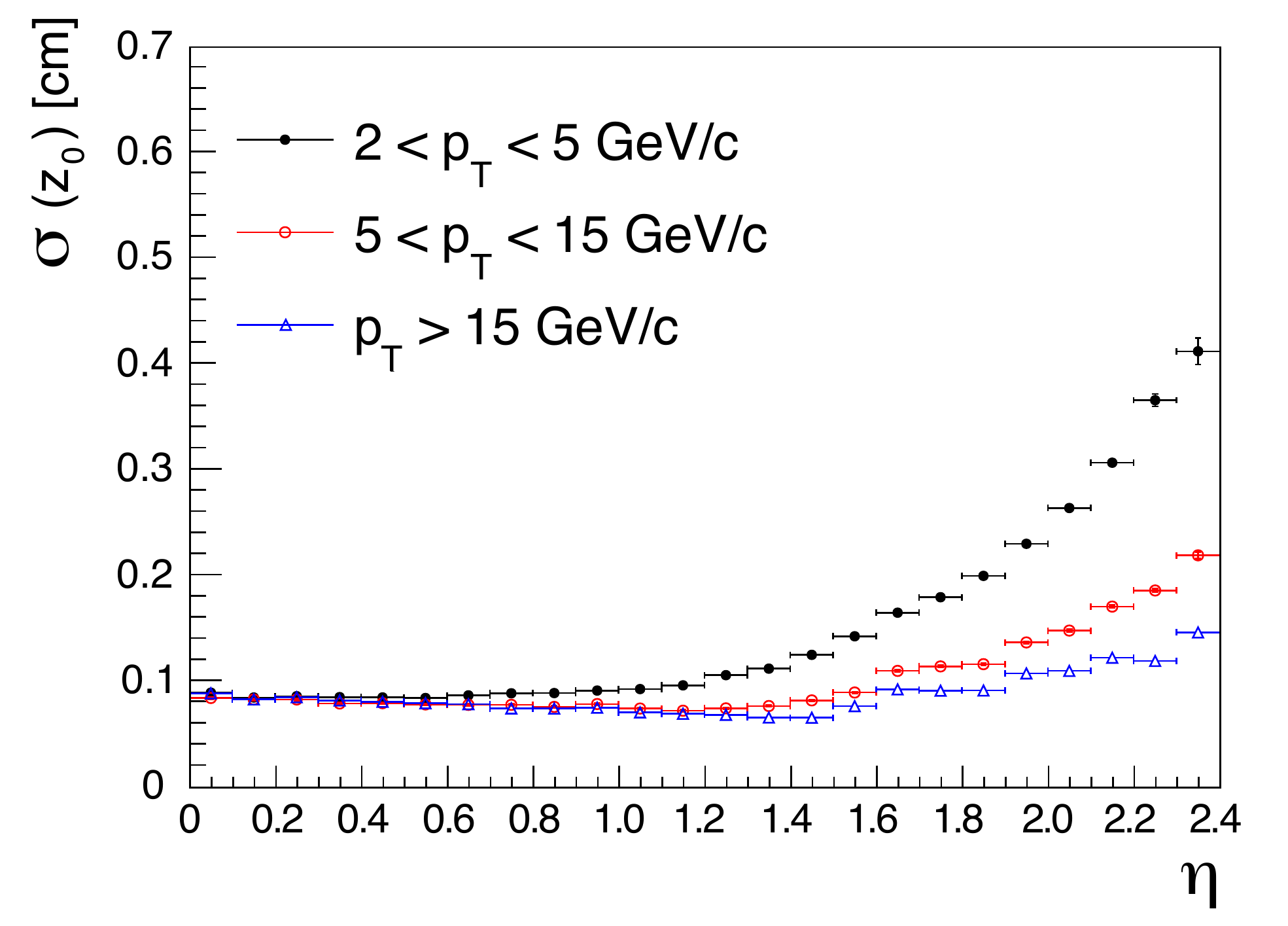}
\includegraphics[width=0.48\textwidth,height=4.8cm]{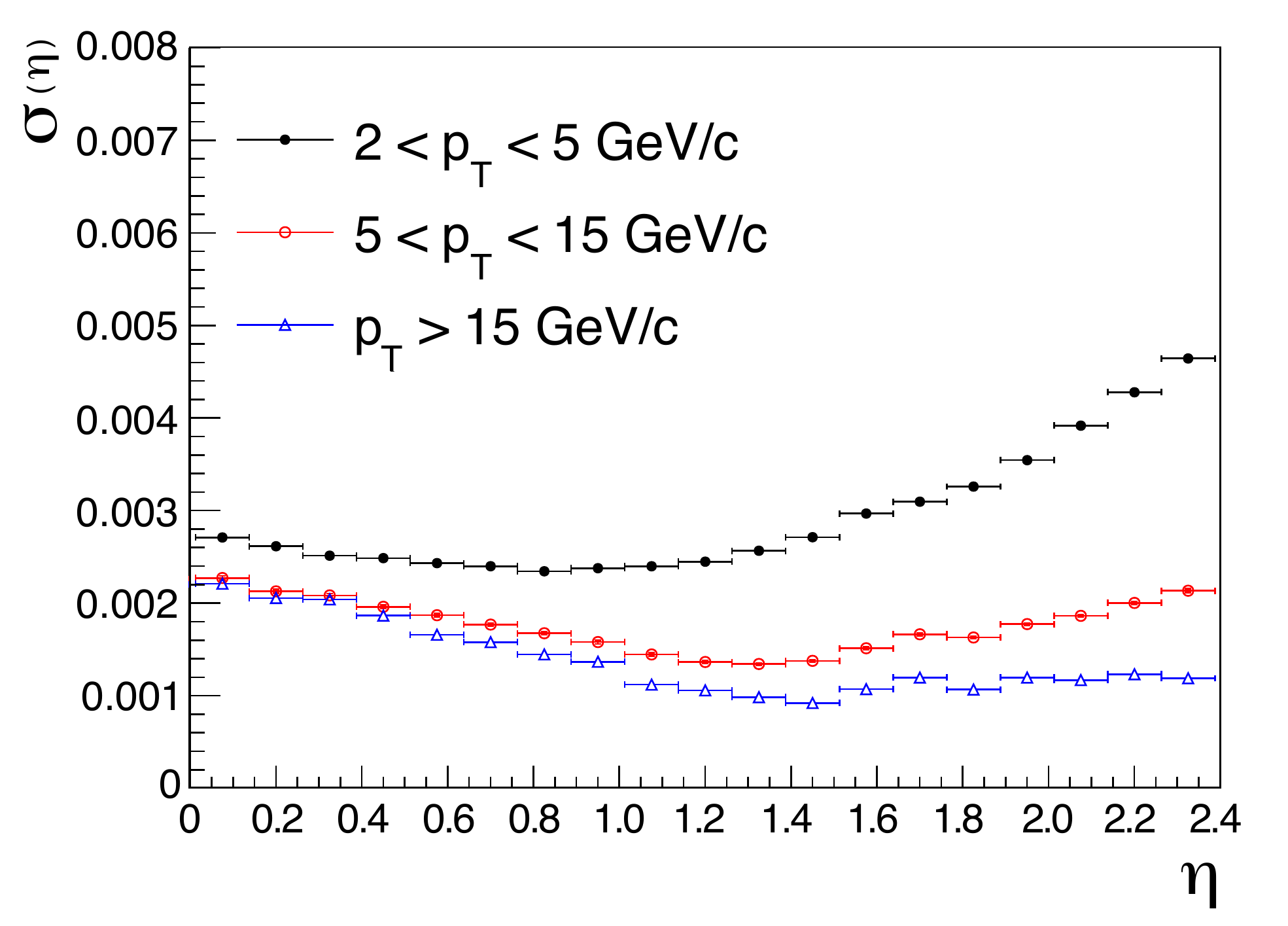}
\caption{Track parameter resolutions for single muon events.}
\label{fig:resolutions}
\end{center}
\end{figure}

An independent sample of muons generated evenly across $\eta$, $\phi$, and $1/p_{\textrm{T}}$ was used to determine the performance of the fit. Figure \ref{fig:resolutions} shows the fitted resolutions of the track $p_{\textrm{T}}$, $\phi_0$, $z_0$, and $\eta$ as a function of true track $\eta$ and for three ranges of true track $p_{\textrm{T}}$. 
In all cases, the obtained resolutions compare very well with the performance of offline simulations and are thus excellent for a track trigger application.  The worsening of resolutions as a function of $\eta$ also follows closely the offline simulation~\cite{CMS-TrackerTDR}.

The relative $p_{\textrm{T}}$ resolution is about 0.5\% for a $2<p_{\textrm{T}}<5\,$GeV track in the region of $|\eta|<0.2$. It may be interesting to notice that if Eq.~\ref{eq:actualcorPhi} is truncated to the first order the relative $p_{\textrm{T}}$ resolution will jump to about 1.6\%. The improvement of the relative resolution with a second order correction is the net effect of addressing the nonlinearities stemming from the functional form of the track equations.

\begin{figure}[!ht]
\begin{center}
\includegraphics[width=\columnwidth]{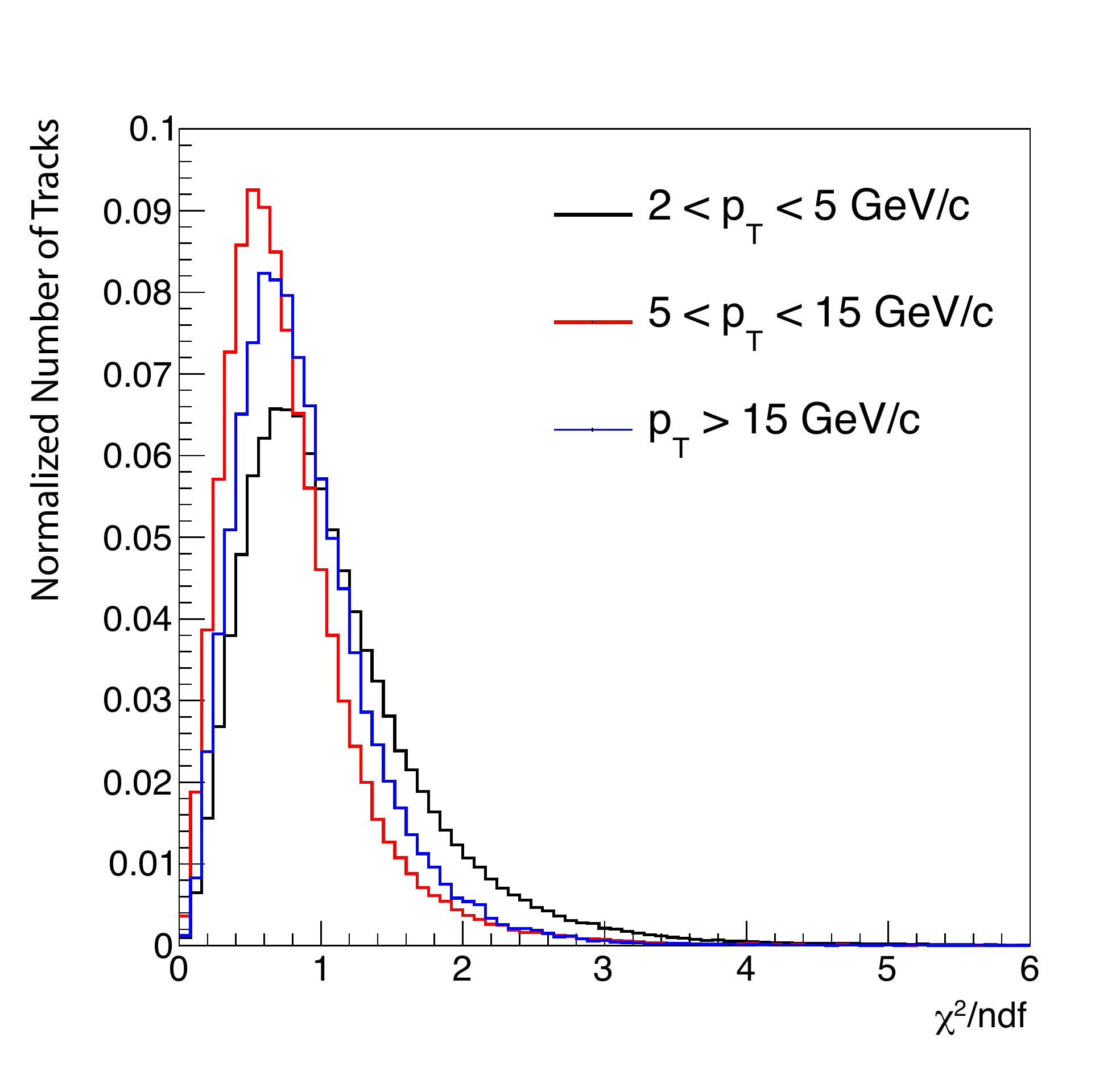}
\caption{Fitted values for track $\chi^2$ for single muon events.}
\label{fig:chi2}
\end{center}
\end{figure}

No significant bias is observed in the fits with the mean central value within the resolution of the true central value for all parameters. The distribution of the reduced $\chi^2$ values is shown in Fig.~\ref{fig:chi2} and is observed to be nicely peaked at values near one, and thus appropriate for a selection requirement of high quality track fits.

% \textcolor{red}{Should we also show the resolutions directly from the pre-estimates to show how much our technique improved things from what you get with a single set of constants?} Found some but they are not in the same format as the ones above, so I guess we skip for now.


% -----------------------------------------------------------------------------
\section{Memory Requirements}
\label{sec:mem}

The number of constants needed to estimate the parameters of a track in a given region of the detector is an important consideration in the hardware deployment of linearized track fitters.

As described in Sect.~\ref{sec:corrections}, our technique begins by computing an initial pre-estimate to obtain coarse track $p_{\textrm{T}}$, $\cot\theta$, and $\tan\theta$ parameters using Eq.~\ref{eq:LTF}. 
The $p_{\textrm{T}}$ is computed using the $\phi$ coordinates of the hits in the six layers and requires only thirteen coefficients (six $A_{\textrm{j}}$, six $\overline{\phi}_{\textrm{j}}$, and one $\overline{p}_{\textrm{T}}$). Following Eq.~\ref{eq:LTF} the over-barred quantities $\overline{\phi}_{\textrm{j}}$ and $\overline{p}_{\textrm{T}}$ represent the average $\phi$ hit positions and track $p_{\textrm{T}}$ parameters about which the linear expansion is performed.

To achieve sufficient precision for the $\cot\theta$ and $\tan\theta$ pre-estimates we utilize both the $z$ and $R$ hit-coordinates using a total of 25 coefficients per pre-estimate (twelve $A_{\textrm{j}}$, six $\overline{z}_{\textrm{j}}$, six $\overline{R}_{\textrm{j}}$, and $\overline{\cot\theta}$). The total number of constants to estimate the pre-estimates for the 14 regions of the detector is then 882. 
% 882 = (13+25*2)*14

This is then followed by the computation of the $\phi$ and $z$ positions of the hits at their projected locations on the idealized layers. This step requires no pre-stored constants except for the radius of the idealized layers. This requires at most six ideal layers per region for a total of $84$ constants. 

Because the pre-estimate returns values of $q/p_{\textrm{T}}$ and $\cot(\theta)$ the transformations of the hit position in Eqs.~\ref{eq:actualcorPhi} and \ref{eq:actualcorZ} require only additions and multiplications and are thus particularly suited for fast applications in FPGAs. The only exception is the $1/R^2$ term in Eq.~\ref{eq:actualNonRadial} that is computed via a lookup table of $16$ constants for the whole detector. 
%This steps takes 15 clock cycles.

The final fit is then performed on the transformed hit positions. Since all remaining nonlinearities are negligible and the $R$ coordinates are fixed, all track parameters and $\chi^2$ components are obtained with either the six $\phi$ or six $z$ transformed coordinates.  In the transverse plane the computation of $p_{\textrm{T}}$, $\phi_0$, and the transverse $\chi^2$ coefficients could be performed with a single set of a $6\times6$ matrix of coefficients, six $\overline{\phi}_{\textrm{j}}$ and a value for $\overline{p_{\textrm{T}}}$ and $\overline{\phi}_{\textrm{0}}$ for a total of 44 constants.  Central values for the $\chi^2$ terms  are zero by construction. However, to improve performance at low transverse momentum, we use two sets of constants, for $p_{\textrm{T}}$ below and above $10\,$GeV, and dynamically choose one based on the $1/p_{\textrm{T}}$ pre-estimate. This step then requires $88 $ constants per region for a total of $1232$ constants.

In the longitudinal plane the computations of the $z_0$, $\cot\theta$, and transverse $\chi^2$ coefficients use another $6\times6$ matrix plus six $\overline{z}_{\textrm{j}}$, $\overline{z_0}$ and $\overline{\cot\theta}$ for a total of 44 constants per region and a total of $616$ constants for all regions. 

Notice that because the transformation restores cylindrical symmetry to the detector we can use a single set of coefficients to estimate the final track parameters and $\chi^2$ components for the full $2 \pi$ azimuthal range in each region. This is a distinct advantage over other approaches that consider each possible combination of modules as a separate entity with a dedicated set of coefficients.

The total number of constants used for the full detector is 2830. % (882+84+16+1232+616).
It is expected that some tracks will leave only five hits instead of six. Depending on the particular hit that is missing it is
possible to have six other sets of combinations, which brings the grand total number of constants to 19,810.

In comparison, the number of constants used in the track-trigger system of the ATLAS detector is close to a million~\cite{FTK}, and the
number used in the CDF experiment for their track trigger system was about eight million~\cite{SVTReview}. As described in the next
section, the number of constants to be stored has implications on the firmware implementation of the linearized track fitter. 

% ATLAS FTK's presentation and 
% ATLAS TDR talks about 64 towers total, with 8 TF/tower with 1250 constant per TF for a total of 640,000
% Guido Volpi's talk (below) talks about 64 towers with about 10-20k per tower for a total between 640,000 to 1.28 M constants
% https://www.dropbox.com/s/c4roax6oi6dtbrx/FNAL_volpig_201506.pdf?dl=0
% CDF's SVT
% https://www.sciencedirect.com/science/article/pii/S0168900206021498
% https://ac.els-cdn.com/S0168900203029450/1-s2.0-S0168900203029450-main.pdf?_tid=a949310c-0a8b-11e8-a296-00000aacb361&acdnat=1517845744_94687de7e32528d08249ae6b92d69f98

% CDF's SVT papers. 8 millions come from the  12 segments in phi* 32 k patterns per segment* 21 constants per pattern.
% [1] https://www.sciencedirect.com/science/article/pii/S0168900206021498
% [2] https://ac.els-cdn.com/S0168900203029450/1-s2.0-S0168900203029450-main.pdf?_tid=a949310c-0a8b-11e8-a296-00000aacb361&acdnat=1517845744_94687de7e32528d08249ae6b92d69f98

% -----------------------------------------------------------------------------
\section{Hardware Implementation and Timing}
\label{sec:firmware}

\begin{figure}[!ht]
\begin{center}
\includegraphics[width=\columnwidth]{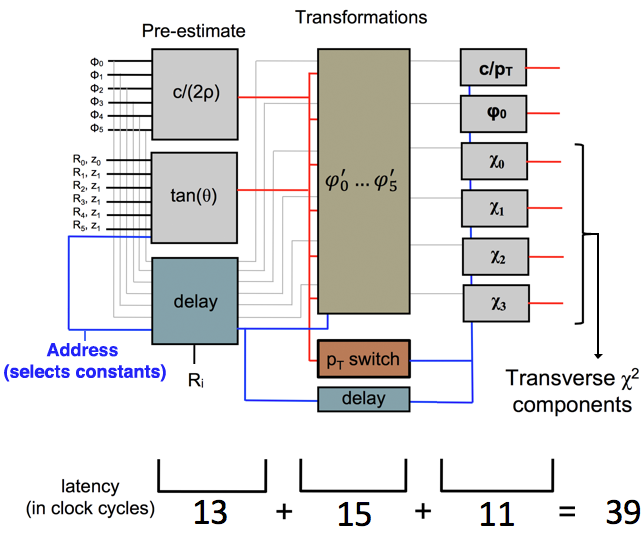}
\caption{Block diagram showing the structure of the firmware implementation of this algorithm for finding the track parameters in the transverse plane. The latency of each step in clock cycles is shown on the bottom.}
\label{fig:fw_diagram}
\end{center}
\end{figure}

The firmware implementation of this technique for the transverse components is shown schematically in Fig.~\ref{fig:fw_diagram}. The algorithm starts by inputting the raw hit positions $(R_{\textrm{j}},\phi_{\textrm{j}},z_{\textrm{j}})$ at the left, which are then used to compute the $q/(2\rho)$ and $\tan\theta$ pre-estimates. These in turn are used to transform the hit positions which are then passed to the linearized track fit to determine the track parameters $q/p_{\textrm{T}}$ and $\phi_{\textrm{0}}$, and the goodness of fit components $\chi_{\textrm{i}}$. A similar schematic is used for the longitudinal components. This firmware is implemented in the Kintex Ultrascale KU060 FPGA. A set of 10,000 tracks was passed through both the firmware and emulator obtaining 100\% bit-to-bit matches.

Within the Kintex Ultrascale class FPGA, the digital signal processing (DSP) structure allows for multiplication of inputs and addition to other inputs, which is a structure that accommodates well the matrix multiplication needed for the linear track fits. The DSPs are arranged in columns and the scalar products are implemented as a set of chained MACC (multiply-accumulate) DSPs, taking advantage of dedicated DSP interconnections within a single column. Because the DSPs are chained they cannot all operate in parallel on the same coordinates and the initial latency is higher than more straightforward configurations. However, the system is fully pipelined and the configuration allows higher clock speeds thus maximizing the throughput and minimizing the overall latency over a large number of consecutive fits.

In addition to track hit positions, each operation also needs access to the relevant constants for the conversion from hit positions to track parameters as in Eq.~\ref{eq:LTF}. The number of needed constants described in Section~\ref{sec:mem} is small enough so that they can be stored in the fabric of the KU060 instead of utilizing the larger block RAM.  Less than 5\% of the total LUT, LUTRAM, and flip-flop resources of the KU060 are needed. This accomplishes three things: it allows the block RAM to be used for other more memory-intensive portions of track finding, it reduces routing delay by placing the constants close to the DSP operators, and it allows a higher clock speed.  The design described here was run at a speed of 500 MHz and utilizes 166 DSPs, or 6\% of the DSPs available on the KU060. The design is fully pipelined and is able to accept a new set of track hits each clock cycle. After the initial fixed latency of 39 clock cycles (78\,ns), a new track fit is returned each clock cycle, thus allowing a total of 461 track fits per $\mu s$ per single instance of the track fitter. Eight instances of the fitter were implemented on a single KU060 while still running at 500\,MHz. 

% -----------------------------------------------------------------------------
\section{Conclusion}
\label{sec:conclusions}

We describe here a novel approach to linearized track fitting based on transformations of the track hit positions to reduce inherent nonlinearities stemming from a combination of detector geometry and the functional form of track trajectories. Unlike conventional techniques that reduce the region of application to contain the effects of the nonlinearities, our approach directly targets and reduces the nonlinearities of the system.  This allows for much better track fit resolutions for a given region of application, or alternatively allows the region of application to be extended significantly for a given desired resolution, thus significantly reducing the required resources to implement the fit.

This approach is developed keeping in mind typical hardware resources and limitations and is demonstrated on a Kintex Ultrascale KU060 FPGA based on the geometry of the CMS upgrade tracking detector. It is shown to have excellent performance in terms of track fit resolutions, hardware latency, and resource utilization requirements. This approach is therefore applicable in the context of very fast track fitting, such as found in hardware triggers for high-energy physics experiments. 

% -----------------------------------------------------------------------------
\section{Acknowledgments}
We thank the CMS collaboration and its Tracker group for allowing the use of the CMS geometry and simulated muon samples in this paper. Both were developed and produced in the framework of the CMS Tracker Upgrade activities. 

This work would not be possible without the funding support of the U.S. Department of Energy, the State of Texas, and the Mitchell Institute for Fundamental Physics and Astronomy at Texas A\&M University.

% -----------------------------------------------------------------------------
\section*{References}

\bibliography{NIMA_Fitter_references}

%% `Elsevier LaTeX' style
%\bibliographystyle{elsarticle-num}
%\bibliographystyle{elsarticle-num-names}
%\bibliographystyle{elsarticle-harv}

\end{document}